\documentclass{appolb}
\usepackage{graphicx}

\begin{document}
\title{Transverse Momentum Fluctuations and Correlations\thanks{Presented by PB at the Workshop Excited QCD,  7-13 May 2017, Sintra, Portugal}
~\thanks{Research supported by National Science Centre grant  2015/17/B/ST2/00101 (PB and SC) and grant 2015/19/B/ST2/00937 (WB)}
}%
\author{Piotr Bo\.zek$^{1}$, Wojciech Broniowski$^{2,3}$, Sandeep Chatterjee$^{1}$
\address{$^{1}$AGH University of Science and Technology,
Faculty of Physics and Applied Computer Science,
al. Mickiewicza 30, 30-059 Krakow, Poland}
\address{$^{2}$The H. Niewodnicza\'nski Institute of Nuclear Physics, Polish Academy of Sciences, 31-342 Krakow, Poland}
\address{$^{3}$Institute of Physics, Jan Kochanowski University, 25-406 Kielce, Poland}
}
\maketitle
\begin{abstract}
We study  fluctuations and correlations of the average transverse momentum of particles emitted in heavy-ion collisions. Fluctuations of the average transverse momentum are related to event-by-event fluctuations of the size and entropy of the initial source. Hydrodynamic calculations using a Glauber model with quark degrees of freedom reproduce the data. We study correlations of the average transverse momentum in different rapidity bins. We propose a definition of the observable that can be directly related to correlations of the collective flow variables.
\end{abstract}
%\PACS{PACS numbers come here}
  
\section{Introduction}

Matter formed in relativistic heavy-ion collisons undergoes 
a collective expansion. The azimuthal asymmetry of the spectra is  studied both experimentally and theoretically as a measure of the initial conditions in the collision, as well as a probe of the properties of the matter formed in the interaction region. The azimuthal asymmetry of the flow is caused 
by the asymmetric acceleration of the fluid. The pressure gradients in the fireball reflect the asymmetry and the fluctuations of the density in the fireball.

\begin{figure}
\begin{center}
\includegraphics[angle=0,width=0.4 \textwidth]{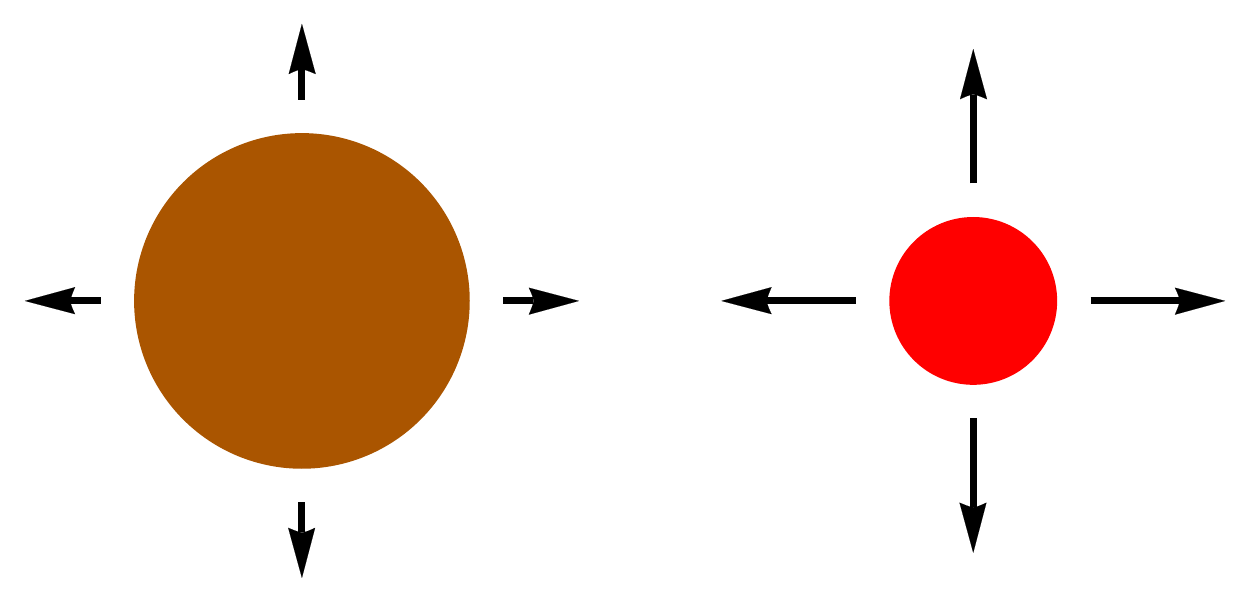}
\end{center}
\vskip -0.5cm
\caption{ Size fluctuations of the initial fireball bring $p_T$ fluctuations in the spectra,  from \cite{Bozek:2017elk}. \label{fig:sizept}} 
\end{figure}  

A very similar mechanism drives the fluctuations of the average transverse flow in the collision  \cite{Broniowski:2009fm}.
 The transverse size of the fireball fluctuates event by event. In events with a small fireball size the accumulated transverses flow is larger than for a fireball with a larger size (Fig. \ref{fig:sizept}).
Transverse momentum fluctuations in Au+Au collisions at $\sqrt{s_{NN}}=200$~GeV
have been calculated in hydrodynamic simulations with event by event
 fluctuating Glauber model initial conditions \cite{Bozek:2012fw} 
and  agree quantitatively with experimental data
 \cite{Adams:2005ka,Adler:2003xq}. 
%The predicted $p_T$ fluctuations 
%slightly overshoot the experimental values. 
This observable may be used as a 
constraint on the magnitude of fluctuations in the initial state.

\section{Transverse momentum fluctuations with wounded quarks}
 
The Glauber Monte Carlo model is often used to obtain the 
  initial state in heavy-ion collisions. 
It contains  event by event fluctuations 
 from random individual nucleon-nucleon collisons. Recently a version of the model involving quark degrees of freedom \cite{Bialas:1977en} has been revived in 
the context of heavy-ion collisons at RHIC and the LHC
 \cite{Adler:2013aqf,Adare:2015bua}. 
The quark Glauber model uses parton degrees of freedom as 
objects that scatter inelastically  in a heavy-ion collision. 
 This model describes fairly well the
 scaling of the multiplicity at central rapidity with the number of 
quarks participating in the collision \cite{Bozek:2016kpf}.

\begin{figure}
\begin{center}
\includegraphics[angle=0,width=0.47 \textwidth]{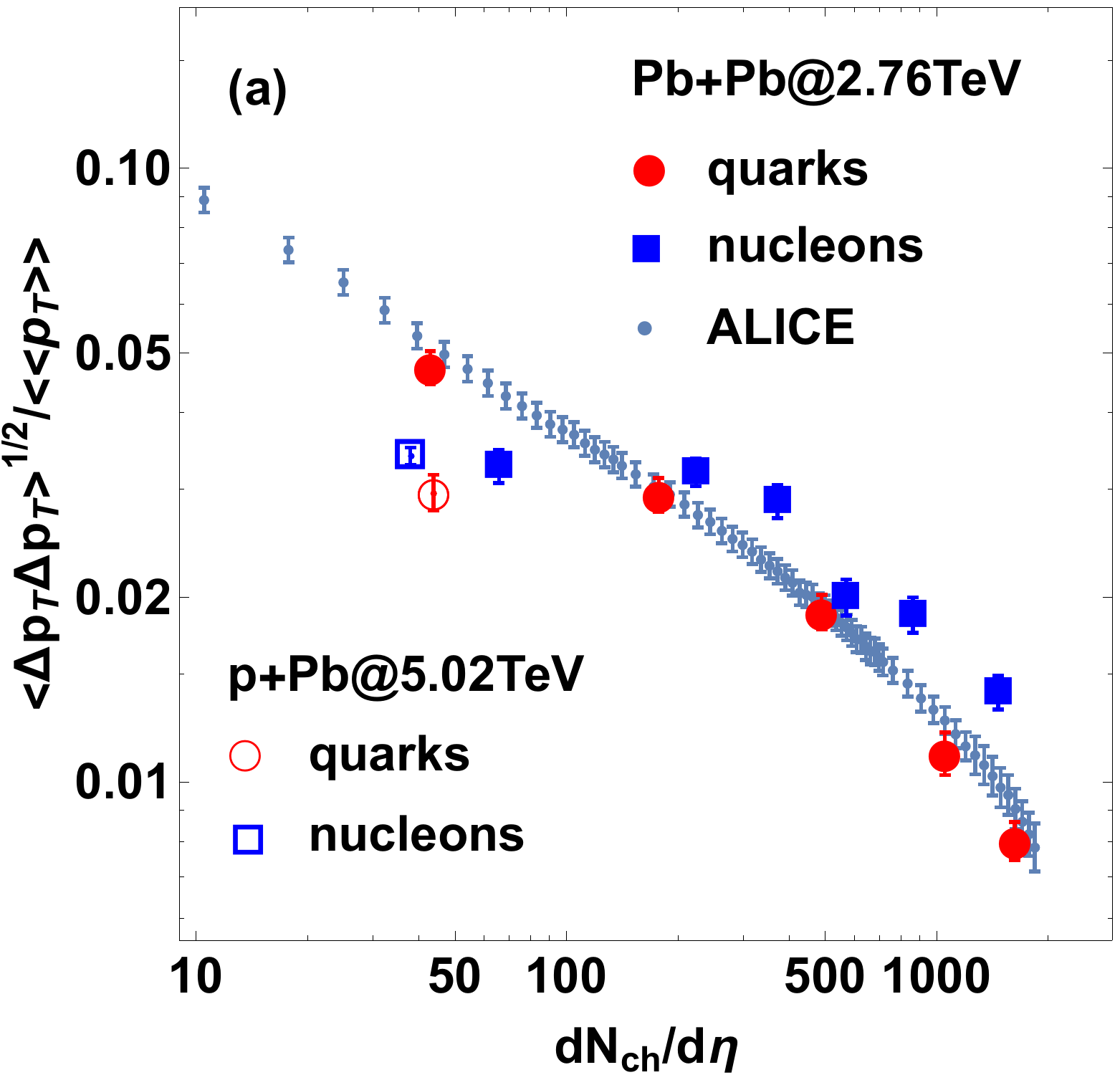}
\end{center}
\vskip -0.5cm
\caption{ The scaled  transverse momentum fluctuations plotted vs charged hadron multiplicity. Results of the hydrodynamic simulations are presented for nucleon Glauber model (squares) and quark Glauber model (circles) initial conditions. The filled symbols correspond to the case of Pb+Pb collisions, whereas the 
empty symbols indicate our predictions for p+Pb collisions at centrality 0-3\%.
The experimental data for Pb+Pb case come from the ALICE Collaboration (small points)~\cite{Abelev:2014ckr}
\label{fig:logdpt},  from \cite{Bozek:2017elk}.} 
\end{figure}  

The variance of the average transverse flow is defined as \cite{Adams:2003uw}
\begin{equation}
C_{p_T}=\langle \frac{1}{N(N-1)}\sum_{i\neq j} (p^i_T -\langle[p_T]\rangle)(p^j_T-
\langle[p_T]\rangle)  \rangle \ ,
\label{eq:cpt}
\end{equation}
where $[\dots]$ denotes the average in a particular event and $\langle \dots \rangle$ the average over the events.
The  above formula, excluding self-correlations, is a robust estimator of 
event by event  fluctuations of the average transverse momentum.  Typically results are presented as the scaled variance
$
\frac{C_{p_T}}{\langle [p_T]\rangle^2}$ \ .

In Fig \ref{fig:logdpt} are presented results from two  viscous hydrodynamic 
calculations, one using nucleon   (squares)
 and one using quark Glauber model of initial conditions (circles).
One notices that the calculation using quark degrees of freedom reproduces the ALICE Collaboration data fairly well. In particular, both the data and the simulation results show a similar multiplicity dependence. We note that this multiplicity dependence deviates from the simple scaling $\frac{C_{p_T}^{1/2}}{\langle [p_T]\rangle} \propto \left(\frac{dN}{d\eta}\right)^{-1/2}$.
Hydrodynamic simulations show that fluctuations of the average transverse
 flow result from the event by event fluctuation of the initial size and entropy of the fireball \cite{Mazeliauskas:2015efa,Bozek:2017elk} on event by event basis.

\section{$p_T$-$p_T$ correlations}
 
Flow generated at two different rapidities is expected 
to be strongly correlated. This correlation results from the early stage of the collision and reflects correlations  in the fluctuations of the 
initial fireball. One can define correlations between any pair of observables related to collective flow or initial entropy: multiplicity, transverse flow,
azimuthal flow coefficients, azimuthal flow angles.

\begin{figure}
\begin{center}
\includegraphics[angle=0,width=0.57 \textwidth]{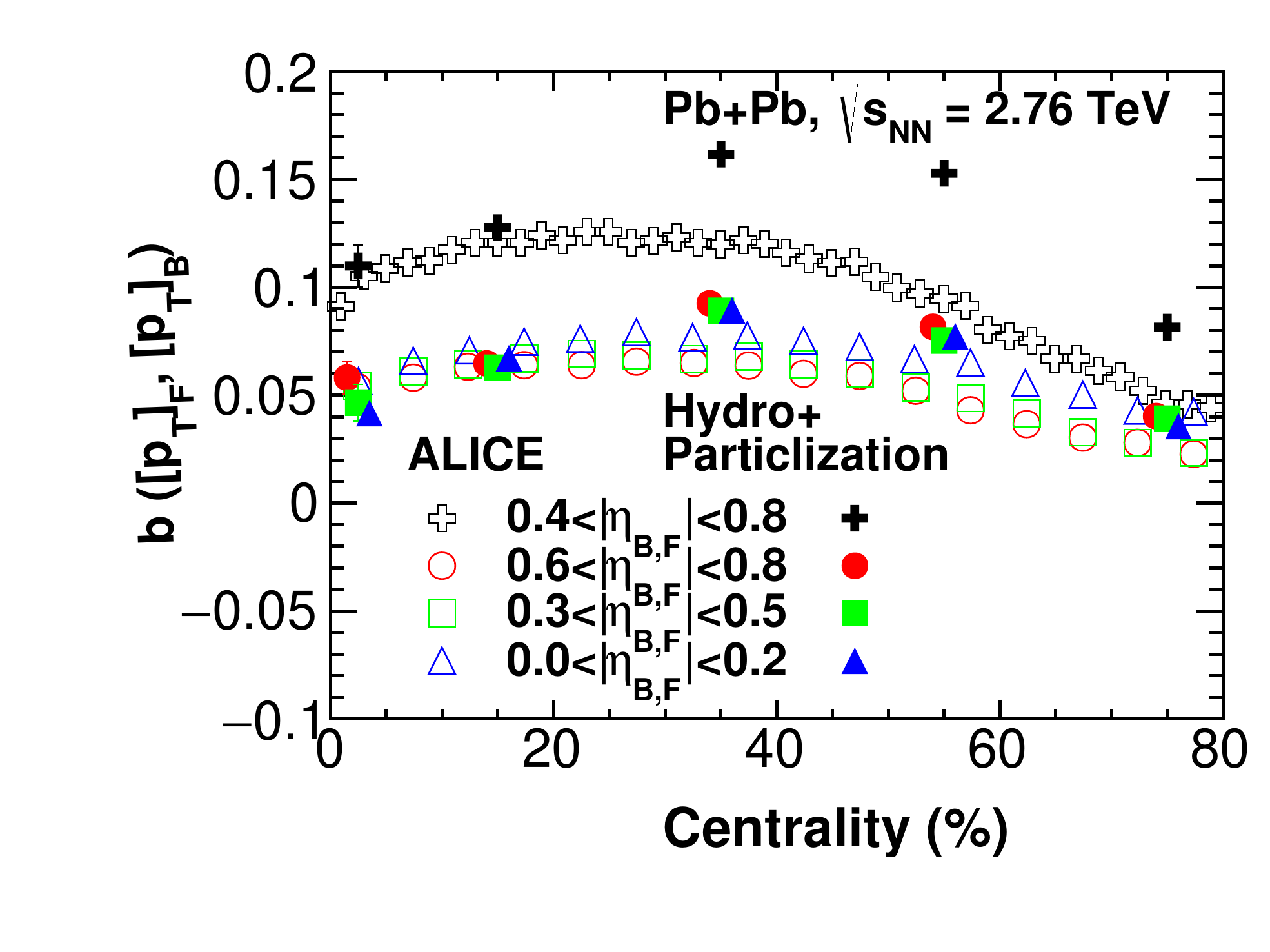}
\end{center}
\vskip -1cm
\caption{  Preliminary data from ALICE Collaboration~\cite{ptposterQM} on 
$b\left( [p_T]_F,[p_T]_B\right)$ are compared to hydrodynamic model predictions,  from \cite{Chatterjee:2017mhc}. \label{fig:balice}} 
\end{figure}  

The correlation of transverse flow in two pseudorapidity regions $F$ and $B$ can be defined using the standard Pearson coefficient of the average transverse 
flow
\begin{equation}
b\left([p_T]_F,[p_T]_B\right)=\frac{\langle ([p_T]_F-\langle [p_T]_F \rangle)  ([p_T]_B-\langle [p_T]_B \rangle) \rangle }{\sqrt{\langle ([p_T]_F-\langle [p_T]_F \rangle)^2 \rangle \langle   ([p_T]_B-\langle [p_T]_B \rangle)^2}} \ .
\label{eq:b}
\end{equation}
The preliminary data for ATLICE Collaboration \cite{ptposterQM} 
are qualitatively reproduced by the model  (Fig. \ref{fig:balice}).
 However,  the main contribution to the denominator 
in Eq. \ref{eq:b} comes from statistical fluctuations in finite 
multiplicity events and not from  fluctuations of the average collective 
flow. Thus, this observable cannot be directly interpreted as a 
measure of the correlation of the collective flow in different  pseudorapidity bins.

\begin{figure}
\begin{center}
\includegraphics[angle=0,width=0.57 \textwidth]{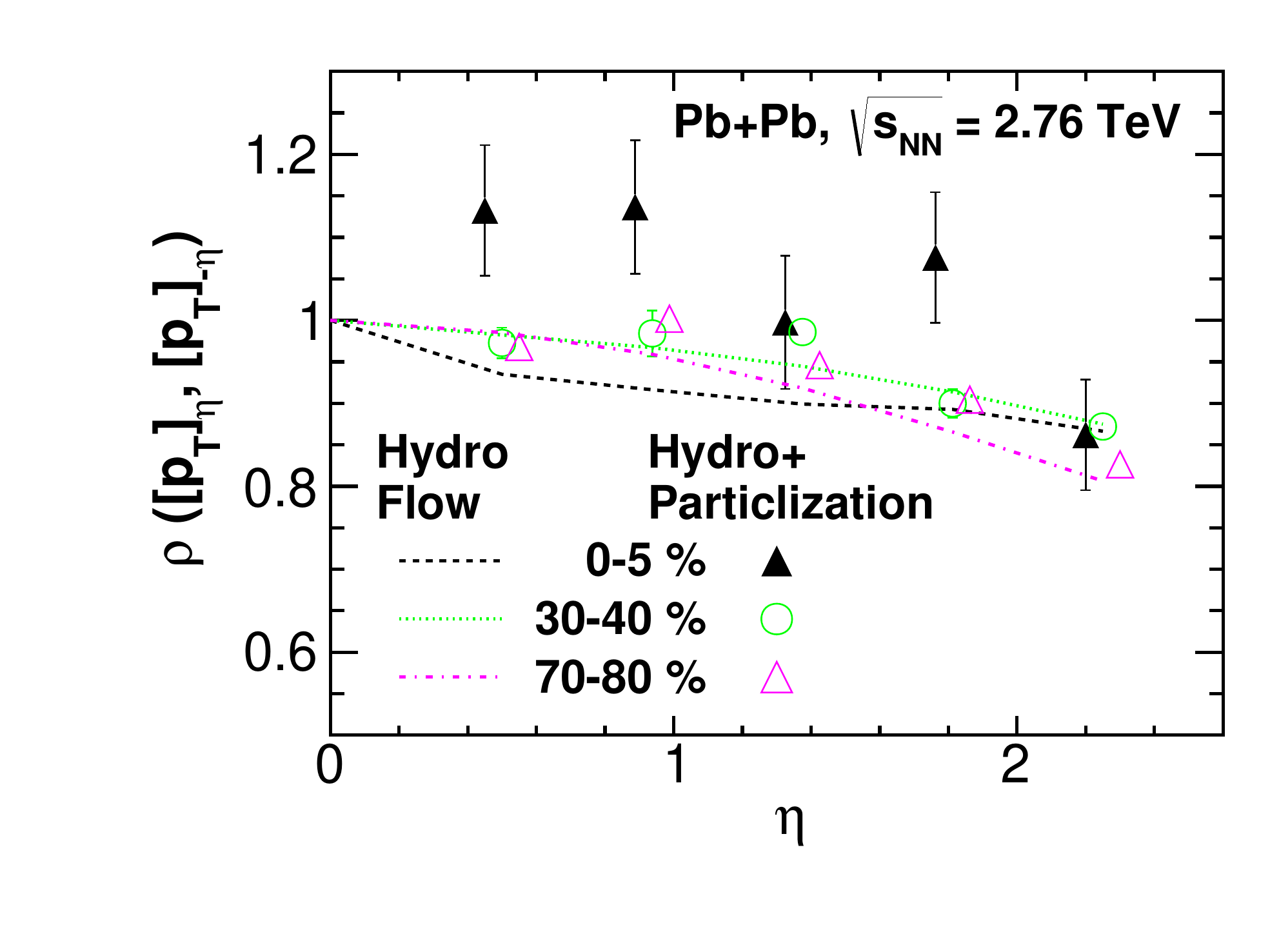}
\end{center}
\vskip -1cm
\caption{  Transverse flow-transverse flow correlation coefficient $\rho\left( [p_T]_\eta,[p_T]_{-\eta}\right)$ as a function of $\eta$. Symbols are for results from realistic finite 
multiplicity events and lines are obtained from spectra integration,
 from \cite{Chatterjee:2017mhc}.  \label{fig:rho}} 
\end{figure}  

The  transverse flow-transverse flow correlation coefficient
can be defined by excluding self-correlations in the definition of the variance \cite{Chatterjee:2017mhc}
\begin{equation}
\rho\left([p_T]_F,[p_T]_B\right)=\frac{\langle ([p_T]_F-\langle [p_T]_F \rangle)  ([p_T]_B-\langle [p_T]_B \rangle) \rangle }{\sqrt{C_{P_T}(F) C_{P_T}(B)}} \ .
\label{eq:rho}
\end{equation}
The results of the calculation using quark Glauber initial 
conditions are shown in Fig. \ref{fig:rho}. We predict that the $p_T$-$p_T$ 
correlation coefficient is close to one ($\rho>0.8$), significantly larger than the Pearson correlation coefficient $b\left([p_T]_F,[p_T]_B\right)$. 
 The results show noticeable non-flow contributions.  

\begin{figure}
\begin{center}
\includegraphics[angle=0,width=0.57 \textwidth]{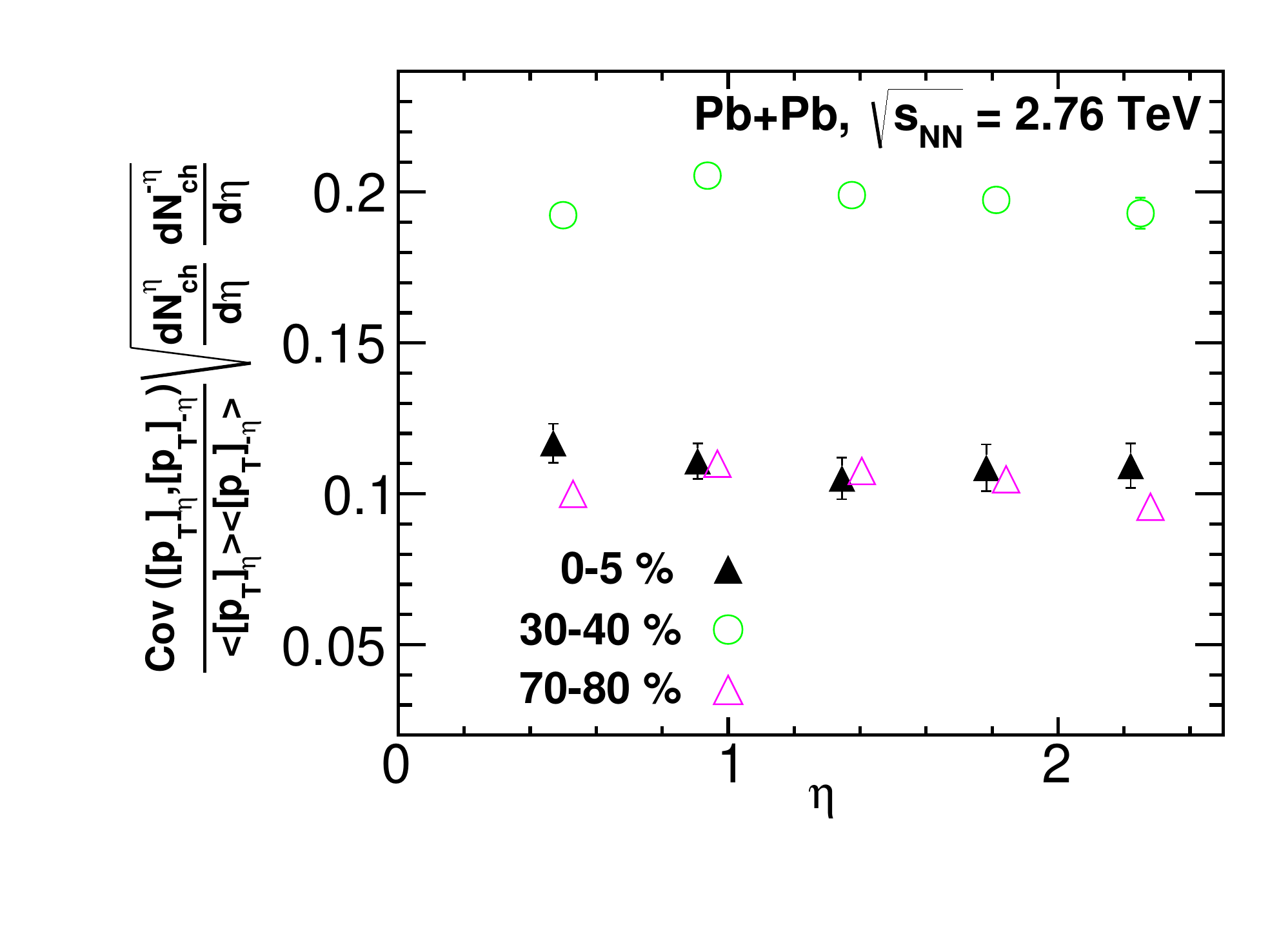}
\end{center}
\vskip -1cm
\caption{   Covariance ${\rm Cov}\left( [p_T]_\eta,[p_T]_{-\eta}\right)$ of the average transverse momentum in two 
pseudorapidity bins scaled by the product of  the square roots of the charged particle densities and by the inverse of the average transverse momenta  in the two pseudorapidity bins. Results are shown as a function of the bin position  $\eta$, for three different centralities.
  \label{fig:covsca}} 
\end{figure}  

Alternatively, the covariance of the average transverse in two 
pseudorapidity bins  flow can be measured.
In Fig. \ref{fig:covsca} is shown the prediction for the covariance
 scaled by the average transverse  momentum. Multiplying by
 $\left(dN/d\eta\right)^{1/2}$ compensates for the trivial multiplicity scaling of fluctuations. The quantity weakly depends on the bin separation, but has an 
interesting centrality dependence, with a maximum for semi central collisions.

\begin{figure}
\begin{center}
\includegraphics[angle=0,width=0.57 \textwidth]{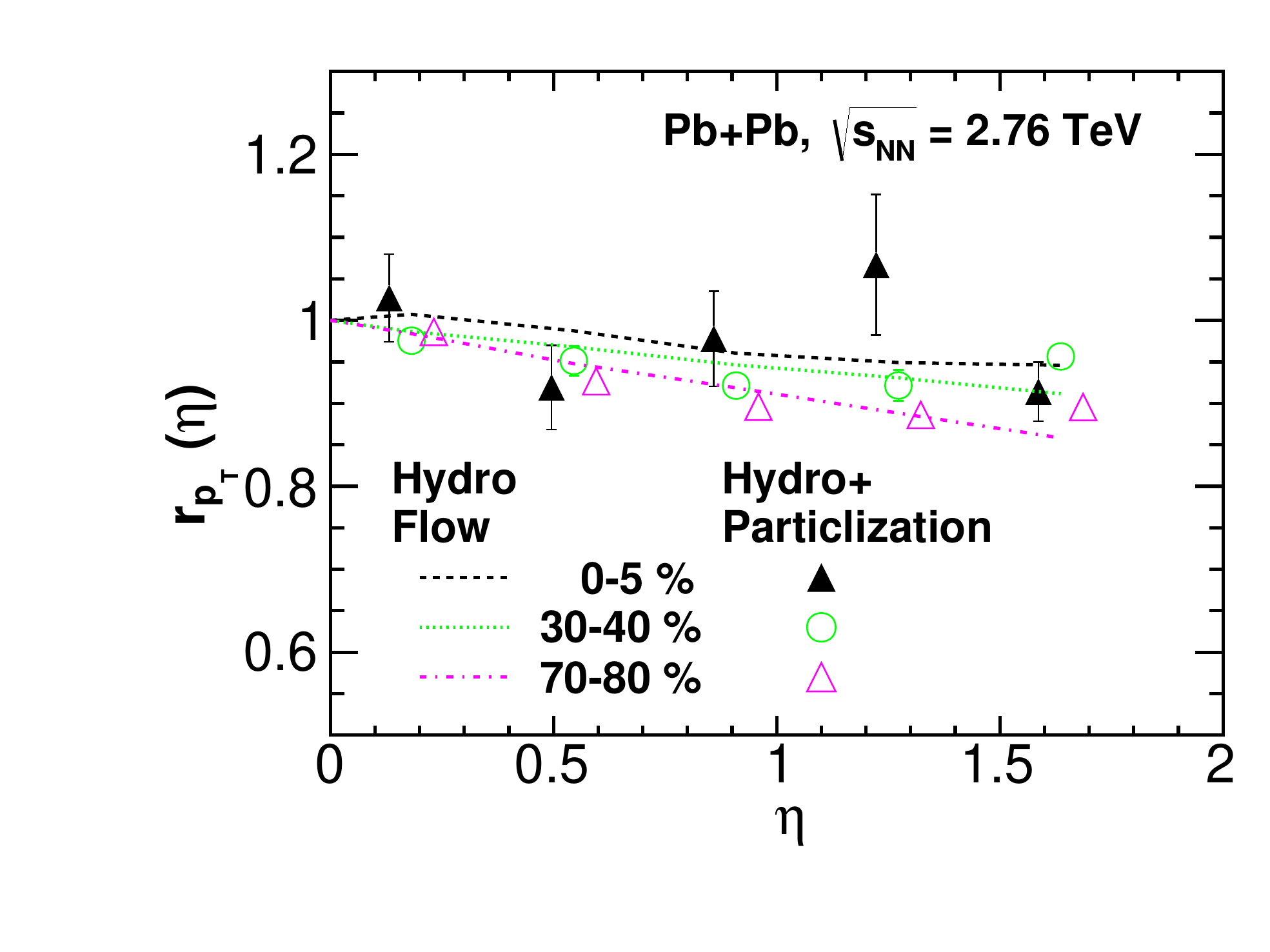}
\end{center}
\vskip -1cm
\caption{   Factorization breaking coefficient ${\rm r}_{p_T}\left(\eta\right)$ of the average transverse momentum at different pseudorapidities.  Symbols are for results from realistic finite 
multiplicity events and lines are obtained from spectra integration.
  \label{fig:rpt}} 
\end{figure}  
%uncomment the following lines to place a figure
%\begin{figure}[htb]
%\centerline{%
%\includegraphics[width=12.5cm]{Fig1}}
%\caption{Plot of ...}
%\label{Fig:F2H}
%\end{figure}

The decorrelation as a function of  pseudorapidity separation can be studied
using a three-bin observable, as proposed by the CMS Collaboration to measure
  azimuthal flow decorrelations   \cite{Khachatryan:2015oea}
\begin{equation}
r_{p_T}(\eta)=\frac{\langle ([p_T]_{\eta_{ref}}-\langle [p_T]_{\eta_{ref}} \rangle)  ([p_T]_{-\eta}-\langle [p_T]_{-\eta} \rangle) \rangle }{\langle ([p_T]_{\eta_{ref}}-\langle [p_T]_{\eta_{ref}} \rangle)  ([p_T]_{\eta}-\langle [p_T]_{\eta} \rangle) \rangle } \ .
\end{equation}
The covariance is measured for two bins (at $\eta$ and $-\eta$)
 with respect to a forward bin at $\eta_{ref}$. In some range of $\eta$ 
one can impose a minimal  pseudotrapidity separation, reducing non-flow contributions.  The decorrelation (or factorization breaking)   of the transverse flow is small and increases with the pseudorapidity separation (Fig. \ref{fig:rpt}).

\section{Summary}

We present a study of the average transverse flow of particles emitted in heavy-ion collisions. Our main observations are
\begin{itemize}
\item
 transverse flow fluctuations result from fluctuations of the  size and entropy of the fireball,
\item 
the hydrodynamic model with quark Glauber initial conditions describes fairly well the variance of the average transverse flow,
\item 
the decorrelation of the transverse flow in two pseudorapidity bins 
can be studied using the correlation coefficient.
\end{itemize}

\bibliography{../../hydr}

\end{document}